\documentstyle[preprint,aps,prl]{revtex}
\begin{document}
\draft
\title{Shell model versus liquid drop model for strangelets}
\author{Jes Madsen}
\address{Institute of Physics and Astronomy, Aarhus University,
DK-8000 \AA rhus C, Denmark}
\date{PhysRevD50, 1 September 1994, in press}
\maketitle

\begin{abstract}
An {\it Ansatz\/} for the curvature contribution to the density of
states for massive quarks in a bag is given and shown to reproduce exact
mode-filling calculations.
A mass-formula for spherical lumps of 3-flavor quark matter is
derived self-consistently from an asymptotic expansion within the MIT bag
model,
taking into account bulk, surface, and curvature contributions.
Shell model calculations are performed for a variety of strange
quark masses and bag constants, and the results shown to match nicely
with the asymptotic expansion.
\end{abstract}

\pacs{12.38.Mh, 12.39.Ba, 24.85.+p, 25.75.+r}

Several experiments with relativistic heavy-ion colliders
\cite{sandweiss91}, as well as cosmic ray searches have been performed
or planned to test the possible (meta)stability of strange quark
matter; a phase of 3-flavor quark matter, that could be
the ground state of hadronic matter (thus more stable than iron)
even at zero temperature and pressure.
The first notion of this possibility seems to be due to Bodmer \cite{bodmer71},
and much work, also on the consequences for neutron stars and cosmology,
has been performed since the idea was resurrected by Witten
\cite{witten84}.
(See Ref.~\cite{madsen91} for reviews and references on strange quark
matter).

It is therefore important to
know the signatures to be expected for small lumps of quark matter
(strangelets) composed of up, down and strange quarks.
An important property is the mass (or energy per baryon).
A calculation involving
mode-filling in a spherical MIT-bag \cite{degrand75}
was performed for $ud$-systems
by Vasak, Greiner and Neise \cite{vasak86}, and for 2- and 3-flavor
systems for a few parameter sets by Farhi and Jaffe
\cite{farhi84} and Greiner {\it et al.} \cite{greiner88}.
Recently, Gilson and Jaffe \cite{gilson93} published a thorough
investigation of low-mass strangelets for 4
different combinations of $s$-quark mass and bag constant.
Such calculations, capable of showing shell-effects etc., are rather
tedious, but important for an understanding of decay modes.
For many applications, including generalization to finite temperature, a
global mass-formula analogous to the liquid drop model for nuclei is of
great use.

An investigation of the strangelet mass-formula within the
MIT bag model was performed by Berger and Jaffe \cite{berger87}.
That investigation
included Coulomb corrections and surface tension effects stemming from
the depletion in the surface density of states due to the mass of the
strange quark. Both effects were treated as perturbations added to a
bulk solution with the surface contribution derived from a multiple
reflection expansion.

Recently it was pointed out that another contribution to
the energy, the curvature term, is dominant (and strongly
destabilizing) at baryon numbers below a hundred
\cite{madsen93a}. A mass-formula (liquid drop model) for systems of
massless quarks including the curvature was derived in \cite{madsen93b}
and shown to fit well with shell model calculations.

However, the density of states correction
due to curvature was only known for massless quarks (and for infinite
mass quarks, where the problem corresponds to the Dirichlet boundary
conditions studied by Balian and Bloch \cite{balian70}), whereas the surface
tension is a function of the mass, vanishing for zero quark mass.
Thus low-mass $u$ and $d$ quarks could be consistently included, but
for realistic $s$-quark masses (expected to be in the range of
100--300MeV) neither $m_s=0$ nor $m_s\rightarrow\infty$ are terribly
good approximations. It was argued in \cite{madsen93a} that it seemed as
if the shell model calculations of Ref.~\cite{greiner88}
required roughly 2 massless curvature
contributions plus the surface energy. As will be shown below, the $s$-quark
curvature contribution indeed goes through zero for intermediate
$s$-quark mass.

In the ideal Fermi-gas approximation the energy of a system composed of
quark flavors $i$ is given by
\begin{equation}
E=\sum_i(\Omega_i+N_i\mu_i)+BV
\label{Estrangelet2}
\end{equation}
Here $\Omega_i$, $N_i$ and $\mu_i$ denote thermodynamic potentials,
total number of quarks, and chemical potentials, respectively. $B$ is
the bag constant, $V$ is the bag volume.

In the multiple reflection expansion framework of Balian and Bloch
\cite{balian70}, the
thermodynamical quantities can be derived from a density of states of
the form
\begin{equation}
{{dN_i}\over{dk}}=6 \left\{ {{k^2V}\over{2\pi^2}}+f_S\left({m_i\over
k}\right)kS+f_C\left({m_i\over k}\right)C+ .... \right\} ,
\end{equation}
where area $S=\oint dS$ ($=4\pi R^2$ for a sphere) and curvature
$C=\oint\left({1\over{R_1}}+{1\over{R_2}}\right) dS$ ($=8\pi R$ for a
sphere). Curvature radii are denoted $R_1$ and $R_2$. For a spherical
system $R_1=R_2=R$. The functions $f_S$ and $f_C$ will be discussed
below.

In terms of volume-, surface-,
and curvature-densities, $n_{i,V}$, $n_{i,S}$, and $n_{i,C}$, the
number of quarks of flavor $i$ is
\begin{equation}
N_i=\int_0^{k_{Fi}}{{dN_i}\over{dk}}dk=n_{i,V}V+n_{i,S}S+n_{i,C}C,
\end{equation}
with Fermi momentum $k_{Fi}=(\mu_i^2-m_i^2)^{1/2}=\mu_i(1-\lambda_i^2)^{1/2}$;
$\lambda_i\equiv m_i/\mu_i$.

The corresponding thermodynamical potentials are related by
\begin{equation}
\Omega_i=\Omega_{i,V}V+\Omega_{i,S}S+\Omega_{i,C}C ,
\end{equation}
where $\partial\Omega_i/\partial\mu_i=-N_i$, and
$\partial\Omega_{i,j}/\partial\mu_i=-n_{i,j}$. The volume
terms are given by
\begin{equation}
\Omega_{i,V}=-{{\mu_i^4}\over {4\pi^2}}\left( (1-\lambda_i^2)^{1/2}(1-
{5\over 2}\lambda_i^2)
+{3\over 2}\lambda_i^4\ln{{1+(1-\lambda_i^2)^{1/2}}\over\lambda_i}\right),
\end{equation}
\begin{equation}
n_{i,V}={{\mu_i^3}\over{\pi^2}}(1-\lambda_i^2)^{3/2}.
\end{equation}

The surface contribution from massive quarks is derived from
$f_S(m/k)=-\left\{ 1-(2/\pi)\tan^{-1}(k/m)\right\}/8\pi$ as
\cite{berger87}
\begin{eqnarray}
\Omega_{i,S}&&={3\over{4\pi}}\mu_i^3\left[{{(1-\lambda_i^2)}\over 6}
-{{\lambda_i^2(1-\lambda_i )}\over 3}\right.\cr
&&\left. -{1\over{3\pi}}\left(\tan^{-1}\left[
{{(1-\lambda_i^2)^{1/2}}\over\lambda_i}\right]-2\lambda_i (1-\lambda_i^2)^{1/2}
+\lambda_i^3\ln\left[{{1+(1-\lambda_i^2)^{1/2}}\over\lambda_i}\right]\right)
\right] ;
\end{eqnarray}
\begin{equation}
n_{i,S}=-{3\over{4\pi}}\mu_i^2\left[{{(1-\lambda_i^2)}\over 2}
-{1\over{\pi}}\left(\tan^{-1}\left[
{{(1-\lambda_i^2)^{1/2}}\over\lambda_i}\right]-\lambda_i (1-\lambda_i^2)^{1/2}
\right)\right] .
\end{equation}
For massless quarks $\Omega_{i,S}=n_{i,S}=0$, whereas
$f_C(0)=-1/24\pi^2$ gives \cite{madsen93a,madsen93b}
$\Omega_{i,C}={\mu_i^2}/8\pi^2$;
$n_{i,C}=-{\mu_i}/4\pi^2$.

The curvature terms have never been derived for massive quarks, but as
will be shown in this Letter, the following {\it Ansatz\/} (found from
analogies with the surface term and other known cases) works:
\begin{equation}
f_C\left({m\over k}\right)=
{1\over{12\pi^2}} \left\{ 1-{3\over 2}{k\over m}\left(
{\pi\over 2}-\tan^{-1}{k\over m}\right) \right\} .
\end{equation}
This expression has the right limit for massless quarks
($f_C=-1/24\pi^2$) and for infinite mass, which corresponds to
the Dirichlet boundary conditions studied by Balian and Bloch \cite{balian70}
($f_C=1/12\pi^2$). Furthermore, the expression gives perfect
fits to mode-filling calculations (see the Figures and discussion below).
{}From this {\it Ansatz\/} one derives the following thermodynamical
potential and density:
\begin{equation}
\Omega_{i,C}={{\mu_i^2}\over{8\pi^2}}\left[\lambda_i^2\log{{1+
(1-\lambda_i^2)^{1/2}}\over{\lambda_i}} +{\pi\over{2\lambda_i}}-
{{3\pi\lambda_i}\over{2}}+\pi\lambda_i^2-{1\over\lambda_i}\tan^{-1}
{{(1-\lambda_i^2)^{1/2}}\over{\lambda_i}}\right] ;
\end{equation}
\begin{equation}
n_{i,C}={\mu_i\over{8\pi^2}}\left[(1-\lambda_i^2)^{1/2}-
{{3\pi}\over 2}{{(1-\lambda_i^2)}\over\lambda_i}+{3\over\lambda_i}
\tan^{-1}{{(1-\lambda_i^2)^{1/2}}\over\lambda_i}\right] .
\end{equation}

With these prescriptions the differential of $E(V,S,C,N_i)$ is given by
\begin{equation}
dE=\sum_i\left(\Omega_{i,V}dV+\Omega_{i,S}dS+\Omega_{i,C}dC+\mu_idN_i
\right) +BdV .
\label{dE}
\end{equation}

Minimizing the total energy at fixed $N_i$ by taking $dE=0$ for a sphere
gives the pressure equilibrium constraint
\begin{equation}
B=-\sum_i\Omega_{i,V}-{2\over R}\sum_i\Omega_{i,S}
-{2\over R^2}\sum_i\Omega_{i,C}.
\label{bag}
\end{equation}
Eliminating $B$ from Eq.~(\ref{Estrangelet2}) then gives the energy for
a spherical quark lump as
\begin{equation}
E=\sum_i(N_i\mu_i+{1\over 3}\Omega_{i,S}S+{2\over 3}\Omega_{i,C}C).
\label{energy}
\end{equation}

The importance of the curvature contribution is demonstrated in Figure
\ref{fig1} which shows the energy per baryon for fixed bag constant of a
bag filled with one flavor of quarks, quark mass ranging from 1 to 450
MeV. The shell model results are derived in a similar manner as
described by Gilson and Jaffe \cite{gilson93},
except that no zero-point correction has
been added. One sees that a significant positive curvature energy arises
for low quark mass (where the surface term approaches zero).
For a quark mass around 150 MeV (for the given
choice of $B$) the curvature term is zero, and the surface term alone
reproduces the shell model calculations, whereas for higher $m_i$ the
surface term overshoots, but the shell model results are reproduced when
including the now {\it negative\/} curvature contribution. For high
masses one approaches the Dirichlet boundary conditions, but since quark
masses are always lower than the chemical potential rather than
infinite, the Dirichlet-limit is only a rough approximation, and the
full surface and curvature terms should be used.

Equally nice fits are obtained for other choices of $B$ and $m_s$. Small
changes in the {\it Ansatz\/} for the curvature density of states leads
to clearly visible errors in the fits to the mode-filling calculations,
so even though the curvature term has not been derived from the multiple
reflection expansion (and attempts to do it have been unsuccessful), it
seems fair to conclude, that this is the proper term, and that surface
plus curvature corrections are sufficient (and equally necessary!) to
reproduce the overall behavior of mode-filling calculations.

Figures \ref{fig2} and \ref{fig3} compare shell model and liquid drop model
calculations for three-flavor strangelets for massless $u$ and $d$
quarks, and massive $s$ quarks for a variety of parameters.
Again, the overall behavior of the energy per baryon is fitted perfectly
by the liquid drop results. For low $m_s$, the energy is roughly that of
bulk plus 3 massless curvature contributions. For intermediate $m_s$, 2
massless curvature terms plus the $s$-quark surface term are important,
and for high $s$-quark mass, where no $s$-quarks are present, the energy
per baryon is given by $E/A$ in bulk plus the curvature terms from 2
massless quarks. (The approach of Ref.\ \cite{berger87} gives {\it
constant\/} $E/A$ in the low as well as high $m_s$-limit since only
surface tension is included, and for intermediate values of $m_s$ the
significant contribution of $u$ and $d$ quark curvature is lacking).

The calculations neglected Coulomb energies, the phenomenological
zero-point energy, and the gluon exchange described by the strong
coupling, $\alpha_s$. Coulomb energies can be self-consistently included
(see \cite{madsen93b}), but are at most a few MeV. The zero-point energy
was included in \cite{gilson93} as $-1.84/R$, and reduces the energies for
very small $A$ somewhat. In $E/A$ this term contributes (for {\it
fixed\/} chemical potentials, which is not quite correct for
$A\rightarrow 1$ \cite{crawford93})
like $A^{-4/3}$, compared to curvature ($A^{-2/3}$)
and surface ($A^{-1/3}$). Significant changes occur only for $A<10$,
and can also be accounted for by an extension of the liquid drop model.
There is no easy way to include the strong coupling.

A detailed study of strangelet decay modes (an important issue for
experimental strangelet searches) must ultimately rely on shell
model calculations.
The main scope of this Letter has been to show, that the general
structure of shell model results for multi-quark systems can be explained
by a liquid drop model mass-formula including volume,
surface, and curvature contributions. Since such calculations are much
easier to perform than the shell model calculations, this allows the
overall structure of strangelets to be studied for a range of
parameters, and facilitates generalizations to finite temperature,
studies of phase transitions and mixed phase structures in heavy-ion
physics, cosmology and neutron stars \cite{mardor91}. For a consistent liquid
drop treatment, the full curvature term
for a massive quark, introduced here for the first time, turns out to be
crucial.

\begin{figure}
\caption{Energy per baryon of 1-flavor quark-lumps at fixed bag constant
($B^{1/4}=145$MeV), for quark masses of 1, 150, 300 and 450 MeV (bottom
to top). For
each mass 4 curves are shown: Horizontal lines are the bulk values;
dotted curves bulk plus surface terms; spiky curves are the shell model
calculations, and smooth curves on top of the shell model results are
the liquid drop model results including surface {\it and\/} curvature.
Notice that the dotted curve is almost indistinguishable from the bulk
curve for $m_s=1$MeV, and overlaps with the full calculation for
$m_s=150$MeV.}
\label{fig1}
\end{figure}

\begin{figure}
\caption{Energy per baryon for 3 flavor strangelets for
$B^{1/4}=145$MeV (the lowest possible value for $ud$-quark matter to
remain unstable). Pairs of curves with the same signature show
shell model and liquid drop model calculations for the same $m_s$
ranging from 50 to 300 MeV in steps of 50MeV (bottom to top). For even
higher $m_s$ no $s$-quarks are present, and results look like those for
$m_s=300$MeV.}
\label{fig2}
\end{figure}

\begin{figure}
\caption{As Figure 2, but for $B^{1/4}=165$MeV (the highest possible
value for stable $uds$-quark matter with $m_s=0$). Now the highest $m_s$
that can be distinguished is 350 MeV.}
\label{fig3}
\end{figure}


\begin{references}
\bibitem{crawford93} Crawford, Desai, and Shaw \cite{sandweiss91}
employ a mass-formula with surface and curvature
contributions to $E/A$ of $cb_6A^{-1/3}+cb_7A^{-2/3}$, where $b_6$,
$b_7$ and $c$ are constants. This form is not appropriate for small $A$
in view of the present investigation. The relative weight of the two
terms depends on $B$ and $m_s$, and the $A$-dependence is not as simple
due to changes in $\mu_i$. For constant $\mu_i$ calculated in bulk the
mass-formulae for $B^{1/4}=145$MeV (as in Fig.\ref{fig2}) are
(with $m_s$ running
from 0 to 300 in steps of 50; all energies in MeV): $E/A=829+351A^{-2/3}$;
$835+61A^{-1/3}+277A^{-2/3}$; $852+83A^{-1/3}+241A^{-2/3}$;
$874+77A^{-1/3}+232A^{-2/3}$; $896+53A^{-1/3}+242A^{-2/3}$;
$911+22A^{-1/3}+266A^{-2/3}$; $917+296A^{-2/3}$.
These ``bulk approximations'' are generally good to
better than 2MeV for $A>100$, 5MeV for $A\approx 50$, 10MeV for
$A\approx 10$, and 20MeV for $A\approx 5$, when compared to the complete
liquid drop model results shown in the Figures.
They always undershoot because the actual chemical
potentials increase when $A$ decreases. For massless quarks the results
scale with $B^{1/4}$. For finite $m_s$ no simple scaling applies. For
$B^{1/4}=165$MeV (Fig.\ref{fig3}) one finds
$985+93A^{-1/3}+265A^{-2/3}$ for $m_s=150$, and
$1027+46A^{-1/3}+284A^{-2/3}$ for $m_s=250$.
\end{references}
\end{document}